# Transiting extrasolar planetary candidates in the Galactic bulge

Kailash C. Sahu*, Stefano Casertano*, Howard E. Bond*, Jeff Valenti*, T. Ed Smith*, Dante Minniti¶, Manuela Zoccali¶, Mario Livio*, Nino Panagia*, Nikolai Piskunov#, Thomas M. Brown*, Timothy Brown†, Alvio Renzini‡, R. Michael Rich§, Will Clarkson*, Stephen Lubow*

*Space Telescope Science Institute, 3700 San Martin Drive, Baltimore, MD 21218
†High Altitude Observatory, 3450 Mitchell Lane, Boulder, CO 80307
¶Universidad Catolica de Chile, Av. Vicuña Mackenna 4860, Santiago 22, Chile
#Uppsala University, Box 515, S-751 20 Uppsala, SWEDEN
‡INAF-Osservatorio Astronomico di Padova, Vicolo dell'Osservatorio 5, 35122 Padova, Italy
§University of California at Los Angeles, Los Angeles, CA 90095-1562

**More than 200 extrasolar planets have been discovered around relatively nearby stars, primarily through the Doppler line shifts owing to the reflex motions of their host stars, and more recently through transits of some planets across the face of the host stars. The detection of planets with the shortest known periods, 1.2 to 2.5 days, has mainly resulted from transit surveys which have generally targeted stars more massive than 0.75 $M_\odot$, where $M_\odot$ is the mass of the Sun. Here we report the results from a planetary transit search performed in a rich stellar field towards the Galactic bulge. We discovered 16 candidates with orbital periods between 0.4 and 4.2 days, five of which orbit stars of 0.44 to 0.75 $M_\odot$. In two cases, radial-velocity measurements support the planetary nature of the companions. Five candidates have orbital periods below 1.0 day, constituting a new class of ultra-short-period planets (USPPs), which occur only around stars of less than 0.88 $M_\odot$. This indicates that those orbiting very close to more luminous stars might be evaporatively destroyed, or that jovian planets around lower-mass stars might migrate to smaller radii.**

Radial-velocity (RV) searches[1-3] have led to the unanticipated discovery of Jovian-mass exoplanets with orbital periods of only a few days---"hot Jupiters." RV surveys favor stars bright enough for high-resolution spectroscopy, and have thus generally been limited to nearby stars that are hotter than early K spectral type, although RV studies are now being extended to M dwarfs[4-5]. Photometric monitoring of microlensing events provides another technique that has recently led to the discovery of a 5.5 Earth-mass planet around a low-mass star[6].

The increasingly successful transit surveys are based on detecting the recurrent dimming of the primary star caused by the silhouetted planet transiting across its face. The first transiting planet, HD 209458b, was found through photometric follow-up of a RV detection[7-8]. Subsequently, five of the transiting candidates found by the OGLE survey[9] have been confirmed as having planetary masses through RV measurements[10-14], and four other planets around bright field stars have been found in separate transit surveys[15-18]. These ten RV-confirmed transiting planets have periods between 1.2 and 4.0 days, masses of 0.36 to 1.45 $M_J$ (where $M_J$ is the mass of Jupiter), and radii of 0.7 to 1.4 $R_J$ (where $R_J$ is the radius of Jupiter).

Transits occur only when the planetary orbit is viewed nearly edge-on. Because of the low probability of such an orientation[19], transit surveys must necessarily cover large numbers of stars, either through wide-angle imaging, or deep imaging in smaller fields with high stellar densities. Since planetary transit depths are rarely more than a few percent, high-precision photometry is required in transit searches. To date, the deepest ground-based transit detections have reached V ~ 17, limiting planet discovery to stars more massive than ~0.75 $M_\odot$, lying within ~2 kpc of the Sun.

The Sagittarius Window Eclipsing Extrasolar Planet Search (SWEEPS) project reported here was conceived as a transit survey that would fully exploit the high spatial resolution and high photometric precision capabilities of the Hubble Space Telescope (HST). We used HST's Advanced Camera for Surveys (ACS) and chose a rich field, lying in the Sagittarius window of the Galactic bulge, which we monitored for periodic small dimmings over a continuous 7-day period.

We show below that at least 7 of our 16 candidates are likely to be genuine planets, rather than brown dwarfs or low-mass stars. Because of the faintness and crowding of the targets, RV confirmation is currently not feasible for the great majority of our targets. However, for two of our brightest candidates, we have used RVs to place limits on the mass of the transiting objects, ruling out stellar companions.

**Stellar properties in the SWEEPS field**

The SWEEPS field lies in the Sagittarius I Window of the Galactic bulge, where the stellar population has a wide variety of metallicities, ranging over -1.5 < [Fe/H] < +0.5[20-21]. Our HST search can detect planetary transits of stars as faint as V ~ 26, which at the distance of the Galactic bulge corresponds to a main-sequence mass of ~0.45 $M_\odot$.

We monitored the SWEEPS field over the 7-day interval 2004 February 22-29, in the V (F606W) and I (F814W) filters. Combination of all of the exposures produces extremely deep V and I images, in which 245,000 stars are detected to V ~ 30 (see Methods for calibration details). The color-magnitude diagram (CMD), presented in Figure 1, shows two components: a dominant population of old stars with a main-sequence turnoff near V=20 exhibiting well-populated subgiant and giant branches, and a much less numerous, closer, and younger population with an unevolved main sequence. We associate the old population with the Galactic bulge, and the younger objects with the foreground disk population, the latter being dominated by the Sagittarius spiral arm[22-23].

The magnitude and color of each star can be used to estimate its mass, temperature, and radius[24-26]. The bulge component is well represented by a population with a distance modulus of $(m-M)_0$ = 14.3, a reddening of E(B-V)=0.64, and a mean chemical composition of [Fe/H]=0.0 and α–element enhancement of [α/Fe]=0.3, similar to values found earlier[27,28]. We adopt a nominal age of 10 Gyr, the isochrone for which is shown as a red solid line in Fig. 1. The blue dashed line shows a representative foreground disk population with an unevolved main sequence at a distance modulus of $(m-M)_0$ = 13.1 and the same reddening as the bulge. Both populations have a significant dispersion in magnitude, due primarily to their depth in the line-of-sight, and also to the metallicity dispersion in the bulge.

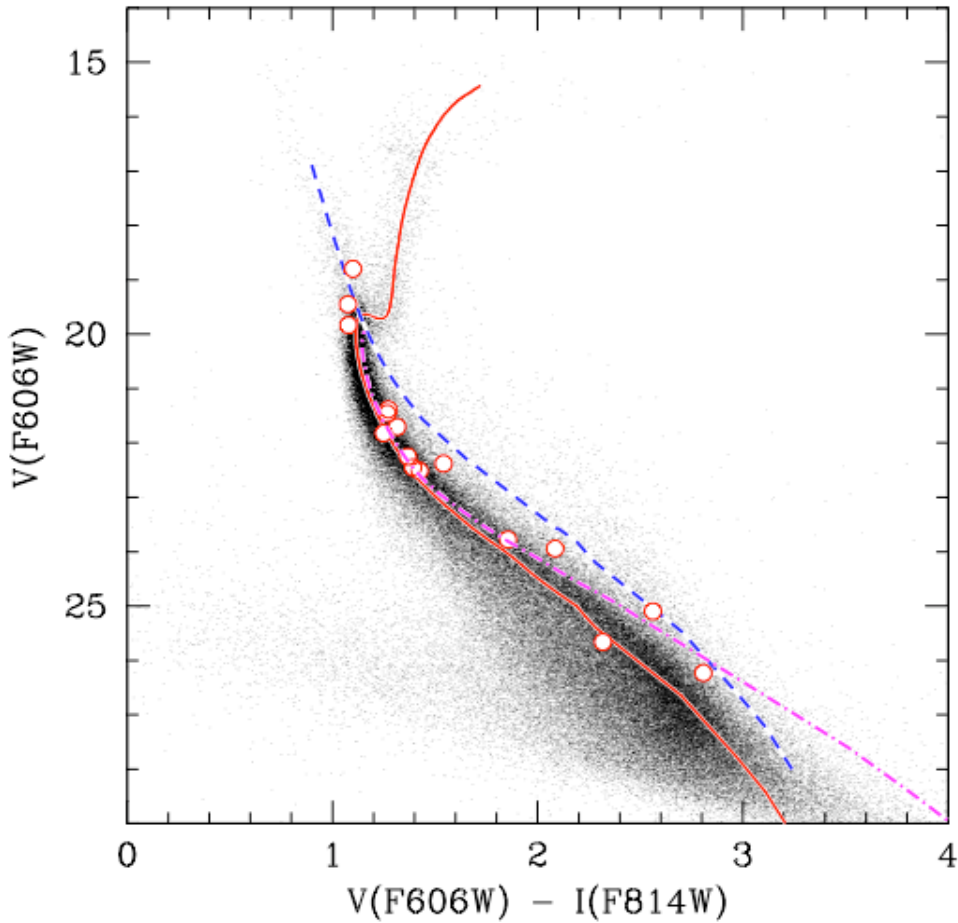

**Figure 1**: |**CMD of the SWEEPS 202″×202″ field in the Galactic bulge.** The CMD of the 245,000 stars was derived from the deep, combined ACS images, with total integration times of 86,106 and 89,835 s in the F606W (V) and F814W (I) filters, respectively. A solar-metallicity isochrone with an age of 10 Gyr (red solid line) fits the dominant bulge population; an unevolved main sequence (blue dashed line) representative of the foreground young disk population is also shown. An estimated isochrone for [Fe/H]=0.5 is shown by the dashed magenta curve. Large circles mark the 16 host stars with transiting planetary candidates.

**Detection and screening of candidates**

In the SWEEPS field there are 180,000 dwarf stars brighter than V=27. We determined a photometric time sequence for each of them by subtracting the total combined image from each individual frame, and then measuring the residual flux at the location of each star in each difference image. We used the box-fitting least-squares (BLS) algorithm[29] to search for transits by calculating a BLS frequency spectrum for each star's time-series data. This algorithm provides approximate transit parameters (period, duration, zero-phase, and transit depth), from which we estimated the signal-to-noise (S/N) ratio of the possible transit signal and retained only those with S/N > 5. For these initial candidates, we then refined the fitting parameters through a $\chi^2$ minimization procedure, now using a code[30] which calculates a full synthetic model light curve. We retained as genuine candidates only those with a statistical significance sufficiently high that the total number of expected statistical false positives in the entire sample is less than one (see Methods).

The resulting list of systems with transiting companions was then screened to remove binaries in which the companions are likely to be low-mass stars rather than planets. We rejected objects showing any of the following properties: (i) a transit depth implying a companion radius > 1.4 $R_J$ (the radii of known transiting exoplanets range from 0.7 to 1.4 $R_J$; larger objects, even at short orbital separations, are likely to be stars[31,32]); (ii) ellipsoidal light variations, implying tidal distortion of the primary by a companion of stellar mass; and (iii) secondary eclipses and/or different transit depths in V and I, indicating a significant light contribution from the companion. We also eliminated objects in which the photo-center of the transit signal (in the difference image) is offset with respect to that of the uneclipsed star (in the direct image), implying presence of an eclipsing stellar binary system whose light is blended with that of a very nearby third star. As an additional check for stellar eclipses, we doubled the period and re-calculated the transit depths. Candidates with statistically significant differences in transit-depths at phase 0.5 and 1.0 in this doubled period were rejected as probable stellar binaries.

This process of elimination resulted in a final list of 16 candidates. The magnitudes of their host stars range from V=18.8 to 26.2, corresponding to stellar masses of 1.24 to 0.44 $M_\odot$. Figure 2 shows a few examples of the observed transit light curves (see supplementary Figs. S-1 and S-2 for all of the light curves).

The rejected candidates include 165 eclipsing binaries in which the secondary is likely to be a stellar companion (Sahu et al., in prep.), out of which 125 show ellipsoidal variations.

**Are the transiting bodies really planets?**

There are, however, several other astrophysical situations that can potentially produce shallow light-curve dips mimicking planetary transits[33], which we now discuss.

5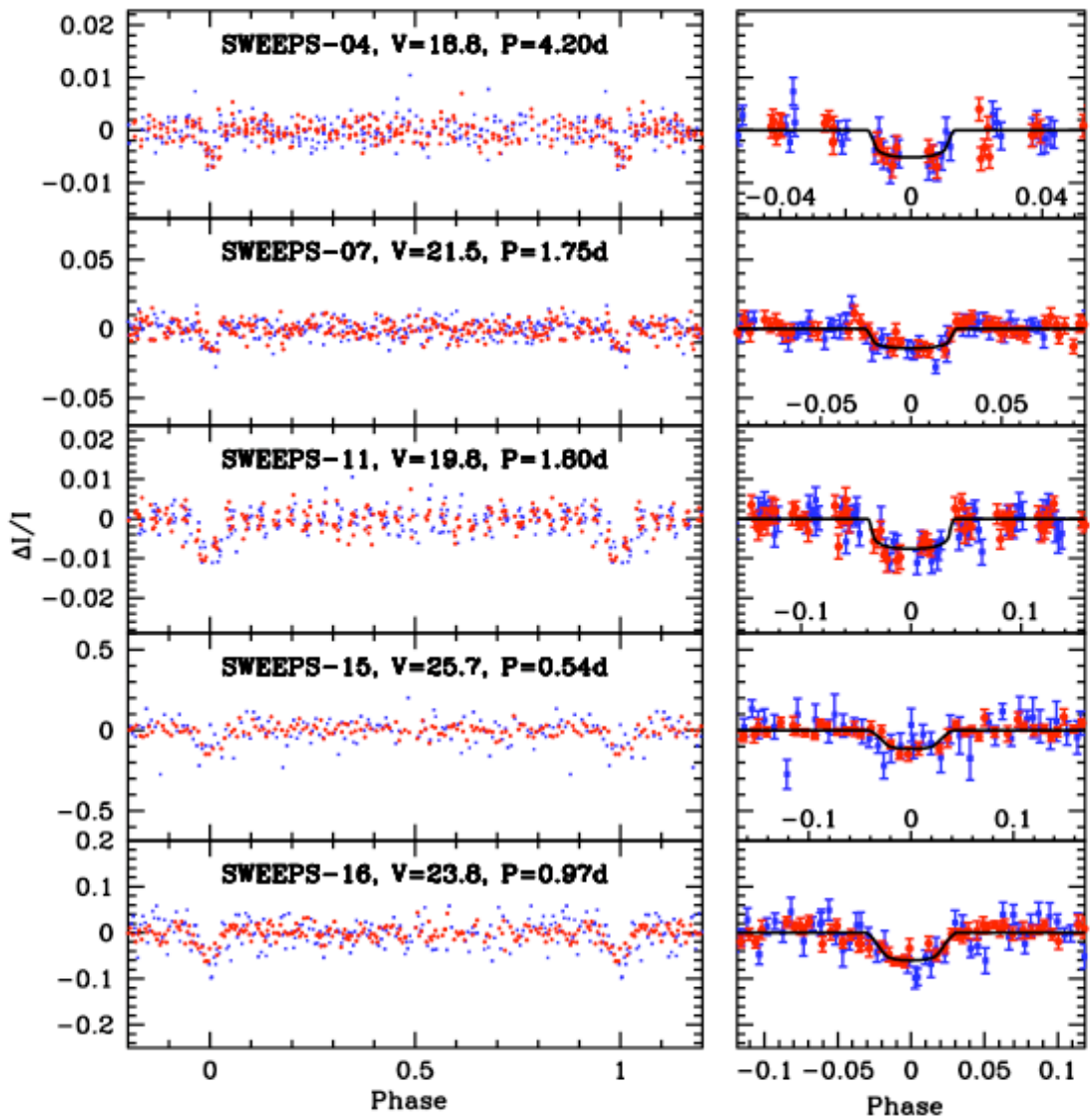

**Figure 2**: | **Example transit light curves.** Five examples of observed light curves which include two with RV measurements, and two USPPs. The left panels show the entire light curve, phased at the derived orbital period, and the right panels show magnified views of the transit with +/-1σ error bars on the individual points. No binning is applied to the data, and the remaining scatter is generally consistent with photon noise. Blue squares are the V-band observations, and red circles are the I-band observations. The black solid curves are the best-fitting model transit light curves. Note that the fainter sources are intrinsically red, so they have significantly higher S/N in the I band.

**TABLE 1: The Transiting Planetary Candidates**

| ID | RA (2000) (h m s) | Dec (2000) (d ' ") | Transit Depth | Per (d) | S/N | Stellar Mass ($M_\odot$) | Stellar Radius ($R_\odot$) | V | I | Planet Radius ($R_J$) | Error* in $R_p$ ($R_J$) | Orbital Radius (au) | Orbital Radius (R*) |
|---|---|---|---|---|---|---|---|---|---|---|---|---|---|
| SWEEPS-01 | 17:58:53.29 | -29:12:33.5 | 0.019 | 1.566 | 8.5 | 0.81 | 0.75 | 22.25 | 20.88 | 1.01 | 0.13 | 0.025 | 7.08 |
| SWEEPS-02 | 17:58:53.38 | -29:12:17.8 | 0.079 | 0.912 | 13.2 | 0.55 | 0.50 | 25.10 | 22.53 | 1.37 | 0.25 | 0.015 | 6.48 |
| SWEEPS-03 | 17:58:53.57 | -29:11:44.1 | 0.015 | 1.279 | 8.3 | 0.79 | 0.72 | 22.51 | 21.09 | 0.87 | 0.11 | 0.021 | 6.35 |
| SWEEPS-04 | 17:58:53.92 | -29:11:20.6 | 0.005 | 4.200 | 8.7 | 1.24 | 1.18 | 18.80 | 17.70 | 0.81 | 0.10 | 0.055 | 9.93 |
| SWEEPS-05 | 17:58:54.60 | -29:11:28.2 | 0.034 | 2.313 | 8.0 | 0.66 | 0.61 | 23.94 | 21.85 | 1.09 | 0.10 | 0.030 | 10.57 |
| SWEEPS-06 | 17:58:57.29 | -29:12:53.4 | 0.004 | 3.039 | 9.5 | 1.09 | 1.36 | 19.45 | 18.37 | 0.82 | 0.21 | 0.042 | 6.68 |
| SWEEPS-07 | 17:58:57.69 | -29:11:14.5 | 0.012 | 1.747 | 10.4 | 0.90 | 0.85 | 21.46 | 20.19 | 0.90 | 0.11 | 0.027 | 6.93 |
| SWEEPS-08 | 17:58:59.24 | -29:13:28.7 | 0.015 | 0.868 | 14.1 | 0.87 | 0.81 | 21.70 | 20.39 | 0.98 | 0.09 | 0.017 | 4.50 |
| SWEEPS-09 | 17:58:59.60 | -29:12:11.8 | 0.020 | 1.617 | 9.5 | 0.79 | 0.73 | 22.45 | 21.06 | 1.01 | 0.12 | 0.025 | 7.38 |
| SWEEPS-10 | 17:59:02.00 | -29:13:23.7 | 0.096 | 0.424 | 11.6 | 0.44 | 0.41 | 26.23 | 23.42 | 1.24 | 0.23 | 0.008 | 4.41 |
| SWEEPS-11 | 17:59:02.67 | -29:11:53.5 | 0.006 | 1.796 | 15.2 | 1.10 | 1.45 | 19.83 | 18.75 | 1.13 | 0.21 | 0.030 | 4.41 |
| SWEEPS-12 | 17:59:04.44 | -29:13:17.1 | 0.014 | 2.952 | 8.7 | 0.86 | 0.80 | 21.82 | 20.57 | 0.91 | 0.11 | 0.038 | 10.33 |
| SWEEPS-13 | 17:59:05.95 | -29:13:05.6 | 0.009 | 1.684 | 7.5 | 0.91 | 0.86 | 21.38 | 20.11 | 0.78 | 0.12 | 0.027 | 6.67 |
| SWEEPS-14 | 17:59:07.56 | -29:10:39.8 | 0.017 | 2.965 | 8.0 | 0.80 | 0.73 | 22.38 | 20.84 | 0.93 | 0.09 | 0.037 | 10.98 |
| SWEEPS-15 | 17:59:07.64 | -29:10:23.7 | 0.099 | 0.541 | 8.5 | 0.49 | 0.45 | 25.66 | 23.34 | 1.37 | 0.30 | 0.010 | 4.90 |
| SWEEPS-16 | 17:59:08.44 | -29:11:40.6 | 0.053 | 0.969 | 15.3 | 0.68 | 0.62 | 23.78 | 21.92 | 1.40 | 0.18 | 0.017 | 5.83 |

*The error in the planet radius ($R_p$) refers to the combined error resulting from the error in the stellar radius (caused mainly by the dispersion in the CMD) and the error due to transit light curve fitting, but does not take into account other possible systematics.

**Grazing stellar eclipses**

A stellar binary with a grazing eclipse can produce a depth similar to that due to a planetary transit. The expected number of grazing incidences can be predicted using the properties of the 40 stellar eclipsing binaries that do not show ellipsoidal variations. By assuming that these are drawn from a population with randomly distributed inclinations, and taking the binary parameters and the detection efficiencies associated with these systems into account, we estimate that a maximum of 1.4 of the 16 candidates could be a grazing system masquerading as a planetary transit.

**Blended eclipsing binaries**

A deeply eclipsing stellar binary, whose light is blended with a brighter constant star, can produce an eclipse in the combined light with a depth similar to that due to a planetary companion of a single star. Blending can occur either because of a chance overlap along the line of sight that causes no detectable shift of the photo-center of the transit signal, or because of membership in a physical triple.

The number of chance overlaps can be estimated from the 165 detected isolated eclipsing stellar binaries down to V=27. Based on their surface density, we expect only 0.3 overlaps with brighter field stars down to V=24, in contrast to the 13 detected planetary candidates that are associated with stars this bright. Assuming that the eclipsing-binary fraction is the same down to V=29, we now expect ~0.8 candidates to be due to chance overlaps in the population down to V=27.

About 10% of all binaries are in physical triple systems, according to statistics in the solar neighborhood[34]. Thus, based on our 165 detected eclipsing binaries, there will be about 17 that are in triples. However, for dilution to produce a shallow transit resembling that of a planet, the third star must lie in a narrow range of magnitudes brighter than the eclipsing pair. Assuming that the third star is drawn from the luminosity function observed in the SWEEPS field, only ~10% of these triples will have transits with depths resembling those due to planets; thus, out of our 16 planetary candidates, ~2 could be due to physical triple systems.

**Radius ambiguity**

An ambiguity arises from the fact that, for masses between about 0.5 and ~150 $M_J$, planets, brown dwarfs, and low-mass stars all have similar radii[31,34,35]. Thus the transit depths that we have measured are insufficient by themselves to distinguish between the three types of companions. However, we provide below some statistical arguments that suggest that a large majority of our transiting companions are in fact likely to be of planetary mass.

To rule out brown dwarfs as significant contaminants of our candidate sample, we note that RV surveys of nearby bright stars show a paucity of brown-dwarf companions in the range 13-80 $M_J$[36]---the so-called "brown-dwarf desert." If the same statistics hold for our bulge stars, the number of brown-dwarf companions will be negligible.

To assess the expected number of low-mass stellar companions in our candidate sample, we refer to results from the RV followup of the OGLE transit candidates. In the OGLE Galactic-bulge and Carina fields, a complete sample of about 60 transiting candidates has been followed up with RV measurements[11-13,37-38]. Most of the screening process we were able to carry out for the SWEEPS transits is not feasible for the OGLE candidate stars, because of their uncertain distances and stellar properties, the lack of multicolor photometry, and the low spatial resolution of ground-based imaging. Nevertheless, the OGLE RV followup has yielded 5 confirmed transiting planetary companions, all of which have radii smaller than the SWEEPS cutoff of 1.4 $R_J$. All of the confirmed OGLE planets have orbital periods less than 4.0 days, and thus occupy the period range to which the SWEEPS survey is sensitive. In this period range, the RV follow-up measurements yielded one confirmed low-mass stellar companion, OGLE-TR-122b, which has a radius of 1.3 $R_J$ (Ref. 35). One more low-mass stellar companion (OGLE-TR-123b) with a radius of 1.2 $R_J$ and a mass of 96 $M_J$ was found in the OGLE followup at a longer period of 7.2 days[38]. From the existing data, we cannot rule out the possibility that there may be a higher proportion of small stellar companions around low-mass dwarfs than around G-K dwarfs. But assuming that similar statistics apply to the Galactic bulge population, we infer that the stellar contamination rate in the SWEEPS sample is likely to be no more than ~29% (two out of seven), and may be half that, because the period of one of the contaminating low-mass OGLE stars with a planetary-sized radius is larger than that of the planets.

**Cumulative sample contamination**

Taking all of the possible contaminants into account, we estimate that at least 45% of our transiting candidates are genuine planets. Considering the fact that no such contamination was found in the 47 Tuc data (see below), this is likley to be a conservative limit. The physical properties of the 16 planetary candidates are listed in Table 1. The table shows that the radii of the SWEEPS planetary candidates range from a sub-Saturnian 0.8 $R_J$ up to our imposed cutoff at 1.4 $R_J$.

## Comparison with globular cluster 47 Tuc

Strikingly different results were obtained in an intensive HST search for transiting planets carried out in 1999 in the globular cluster 47 Tucanae[39]. Some 34,000 cluster stars were monitored throughout an 8-day period. In spite of similar stellar crowding, transit-depth sensitivity, and mass range of the surveyed stars, the population of transiting planetary candidates found in the Galactic bulge is conspicuously absent from the cluster. We emphasize that 11 eclipsing stellar binaries were found in 47 Tuc[40], but not a single transiting object with a radius consistent with a planetary companion.

The lack of planets in 47 Tuc could be either due to its lower metallicity (Fe/H ~20% of solar) or high stellar density. The strong correlation between the planet frequency and stellar metallicity seen in the solar neighborhood[41,42] points towards metallicty as the cause, and provides a plausible explanation for the pronounced difference in the planetary incidence between 47 Tuc and the Galactic bulge.

## Mass limits from RV measurements

Although we argue that most of our transiting candidates are of planetary mass, unambiguous confirmation can come only from RV measurements. These are very challenging with current technology in the SWEEPS field due to the faintness and crowding. We have nonetheless been able to place meaningful limits on the companion masses of two of our brightest candidates, SWEEPS-04 and SWEEPS-11, which lie in relatively uncrowded locations. During 2004 June 22-25, we observed both stars simultaneously, using a fiber-fed mode of the UVES echelle spectrograph at the 8-m Very Large Telescope of the European Southern Observatory at Cerro Paranal, Chile. Two detectors covered the wavelength ranges 4812-5750 Å and 5887-6759 Å. In 31 total hours of exposure time, we attained a peak S/N ratio of 7 to 10 per resolution element in the coadded spectrum of each star.

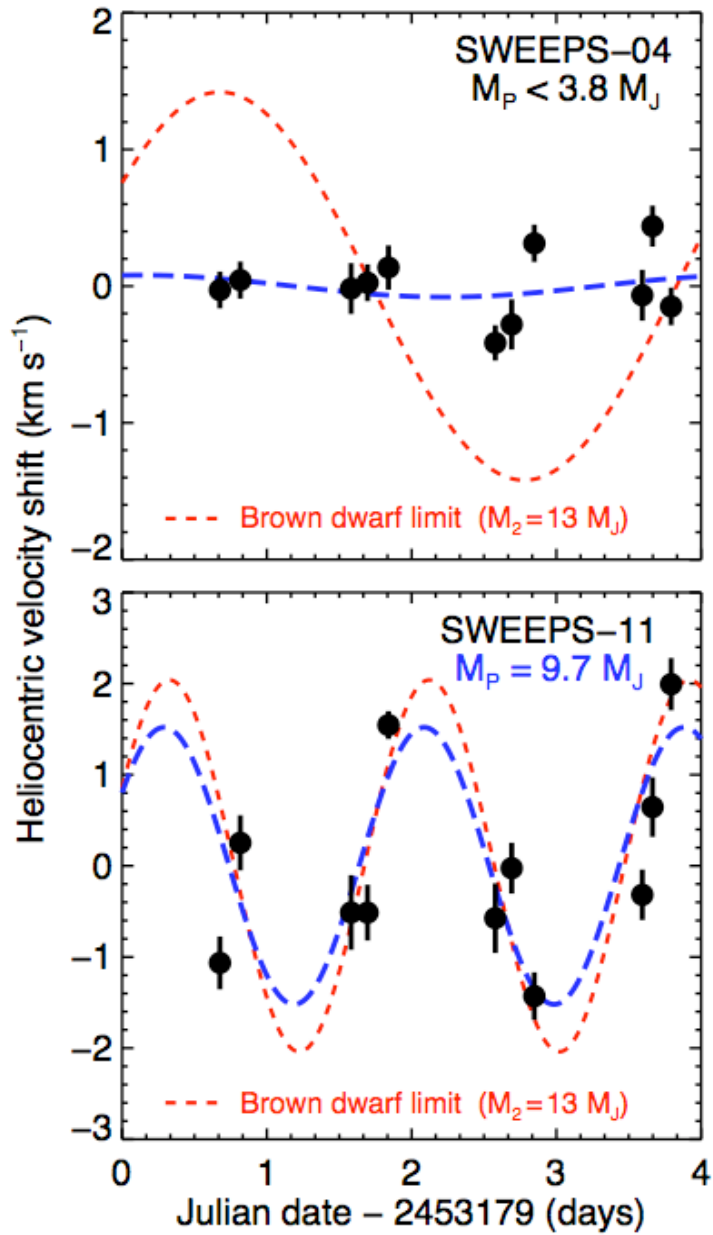

**Figure 3**: | **RV measurements**. RV variations (black dots) for SWEEPS-04 and -11 from VLT spectra. For SWEEPS-04, the best formal fit (blue curve) has insignificant amplitude, and at the 95% and 99.9% confidence levels, we rule out companions more massive than 3.8 $M_J$ and 5.3 $M_J$. For SWEEPS-11, the best formal fit (blue curve) has an amplitude of K=$1.5^{+0.9}_{-0.7}$ km s$^{-1}$ ($M_p$ = $9.7^{+5.6}_{-4.5}$ $M_J$), formally yielding a 99% likelihood of detection, unless residual systematic errors are significant. Expected RV variations for a minimum-mass brown dwarf (13 $M_J$, dashed red curves) are also shown. The photometric and RV phases for SWEEPS-11 are consistent.

We found 51 stellar absorption features strong enough to provide a useful constraint on RV. We fitted each individual feature with a shifted and binned solar spectrum that had been numerically degraded to match our velocity resolution of 5 km/s. We then rejected outliers and calculated a noise-weighted mean RV for 11 separate epochs.

Fig. 3 shows our measured RVs and formal uncertainties for each epoch. For SWEEPS-04, the absence of detected stellar velocity perturbations rules out at the 95% confidence level a companion more massive than 3.8 $M_J$ (K>0.42 km s$^{-1}$). For SWEEPS-11, we probably detect a RV variation with an amplitude of $1.5^{+0.9}_{-0.7}$ km/s, corresponding to a companion mass of $9.7^{+5.6}_{-4.5}$ $M_J$, but a non-detection is also possible if residual systematic errors are significant. Setting aside the low probability of triple systems, the RV data suggest that both of these companions are planets (or a low-mass brown dwarf for SWEEPS-11), but certainly not stars.

**Ultra short-period planets**

Five of our candidates have periods of less than 1.0 day. We call them USPPs, noting that the shortest orbital period yet found among RV-confirmed planets is 1.2 days, whereas our USPPs extend the periods down to 0.42 day. Statistical analysis of possible false positives suggests that at least 2 of these USPPs might be genuine planets.

All 5 USPPs orbit stars of less than 0.88 $M_\odot$. USPPs thus seem to be analogs of hot Jupiters, but around lower-mass stars. We note that USPPs are not expected to be especially hot compared to previously known "hot Jupiters", since the irradiance from the low-mass primary at their locations is comparable to that of planets found around more massive stars. In fact, we argue below that irradiance levels may be one of the reasons why USPPs are not found around more massive stars. We also note that, in units of host stellar radii, the USPPs are no closer to their parents than the closest of the ordinary hot Jupiters. For example, the smallest orbital separation in units of stellar radii among the SWEEPS candidates is that of the USPP SWEEPS-10 (4.412 R*), while the next smallest is that of SEEPS-11, a 1.8 day hot Jupiter orbiting a 1.1 $M_\odot$ star at 4.414 R*. The shortest period RV-confirmed planet, OGLE-TR-56b has an orbital radius of 4.397 R*.

USPPs do not raise an issue of stability against tidal breakup, since even at the closest of the observed separations a planet of more than 1.6 $M_J$ will lie within its Roche radius.

**Physics of the planetary candidates**

The positions of the 16 host stars in the CMD of the SWEEPS field are marked with open circles in Fig. 1. Below V ~ 23, there is an increasing tendency for the hosts to lie redward of the bulge ridge-line. As shown in Fig. 1, the high-metallicity isochrone (dash-dotted magenta curve) also exhibits a progressively increasing shift redward as one moves below V ~ 23. The locations of the primaries found in our sample are thus consistent with the earlier findings that high-metallicity stars are more likely to host planets. (Although high metallicity provides a natural explanation for the redward shift at faint levels, we cannot exclude the alternative that some of the faint hosts belong to the foreground disk. Proper motions of the host stars from future second-epoch observations will resolve this issue.)

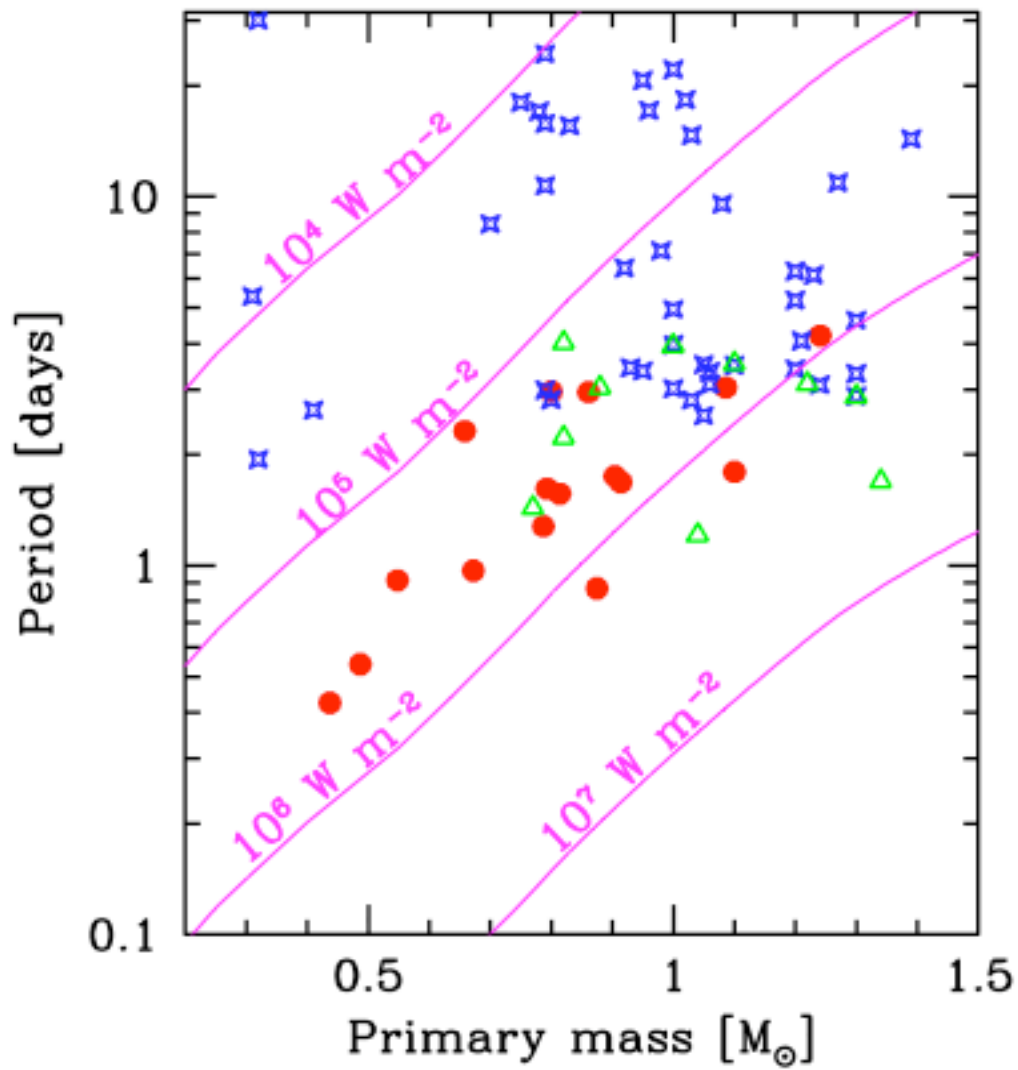

**Figure 4 | Plot of orbital period against host-star mass.** Orbital periods and host-star masses for extrasolar planets. Solid red circles are the 16 new SWEEPS candidates, green triangles are previously discovered transiting planets, and red crosses are lower mass limits for planets detected through RV variability. The SWEEPS candidates extend the range of planetary orbital periods down to 0.42 days. Very few planets have irradiances above $2 \times 10^6$ W/m$^2$ (which corresponds to an equilibrium temperature of 2000 K) and none in the SWEEPS sample. The absence of ultra-short-period planets around stars more massive than 0.9 M$_\odot$ may be due to irradiative evaporation.

The sample of RV-detected planets in the solar neighborhood indicates that the frequency of occurrence of Jovian planets is 5 to 10% for F through K dwarfs, about one-tenth of which are hot Jupiters with periods less than 4.2 days; the frequency of planets around M dwarfs is <5%[4,37]. Our sample of candidates mainly belong to the Galactic bulge, the farthest such sample in the Galaxy, where the metallicity distribution is broader than in the solar neighbourhood[21,22]. However, after taking into account the relation between planet frequency and metallicity in the local sample[41], we would expect the overall planet frequency in the bulge to be similar to that in the solar neighborhood. A major factor that reduces the completeness of transit surveys at longer periods is the lower geometric probability of a transit occurring, which is inversely proportional to the orbital radius[19]. After correcting for geometric transit probability and our detection efficiency (see Methods), we find that our 16 candidates (if all of them are assumed to be genuine planets) imply that about 0.42% of bulge stars more massive than ~0.44 $M_\odot$ are orbited by Jovian planets with periods less than 4.2 days. Due to the small-number statistics and uncertainties in the detection efficiencies, this fraction is uncertain by perhaps a factor of 2. Thus, within the statistical errors, the overall frequency of occurrence of planets derived from the SWEEPS data is consistent with that in the solar neighborhood. For host stars more massive than 0.75 $M_\odot$, the observed period distribution of the SWEEPS planets is also consistent with that in the solar-neighborhood sample. However, for lower-mass stars, the SWEEPS period distribution is systematically shifted to shorter periods.

Figure 4 plots the orbital periods against the host-star masses for the SWEEPS candidates, as well as for planets previously discovered in RV and transit surveys. This figure shows that USPPs occur preferentially around low-mass stars. Thus, the reason that USPPs have not been revealed until now appears to be simply that previous transit and RV surveys have not reached down to sufficient numbers of low-mass, low-luminosity host stars with suitably high metallicity.

According to current models, Jupiter analogs with orbital periods of a few days cannot form in situ, but must instead form at larger radii and then migrate inwards at an early stage via interaction with the circumstellar disk. Migration halts either when the disk dissipates or when the planet approaches the magnetically truncated inner edge of the disk at a few stellar radii[43,44]. The existence of USPPs may imply that Jovian planets around lower-mass stars can migrate to smaller orbital separations, perhaps because the size of the inner disk hole decreases with decreasing stellar radius.

In order to investigate the cause for the absence of USPPs around stars above 0.9 $M_\odot$, we calculated the surface fluxes received at the planets using the known temperatures and radii of the host stars and the orbital separations. The diagonal lines in Figure 4 mark loci of constant irradiation at the planetary surface. There are very few planets found at irradiations larger than 2 x $10^6$ W $m^{-2}$ (corresponding to an equilibrium surface temperature of ~2000 K[45]), and none in the SWEEPS sample. (It has been suggested[46] that the XUV and Lyman-alpha irradiation may be more important than the total irradiation in driving the escape of planetary atmospheres, but the former scales with the total irradiation times a factor related to the stellar Rossby number[47].) Fig. 4 thus suggests that planets that find themselves in such extreme thermal environments are irradiatively evaporated[46,48], on a timescale that is less than ~10 Gyr. In close proximity to less luminous stars, however, these ancient worlds have survived for more than twice the age of our own solar system.

**METHODS**
**Time-series and absolute photometry**

During the 7-day SWEEPS monitoring observations, we obtained 254 and 265 exposures in the V and I bands, respectively. Each HST orbit had a visibility period of 52 minutes, during which we typically obtained 3 exposures each in V (F606W) and I (F814W). Exposure times in each filter were 339 s, giving a photometric accuracy per observation of about 0.003 mag at V=20 and 0.04 mag at V=25. A typical planetary transit lasts 1.5 to 3 hours, so there are generally 5 to 10 individual observations during each transit.

Our approach for time-series photometry is similar to the one followed for the analysis of the 47 Tucanae data[39,49]. The method relies on the formation of difference images, which requires simultaneous determination of image-to-image spatial offsets and modeling of the signal response to spatial offsets and telescope focus changes. The resulting difference images contain noise due to Poissonian photon statistics and CCD readout noise, plus genuine residuals due to stellar variability. Photometric time series are derived from the difference images by performing both aperture and PSF (point spread function)-fitting photometry, and on a star-by-star basis selecting the one that provides the lower noise. The resulting time series are then cleaned of any remaining correlations with x,y position, to provide the final photometry (plus errors) for each target.

The color-magnitude diagram for the SWEEPS field was determined from absolute photometry (Vegamag system) performed on twice-oversampled co-added images of the entire dataset in V and I using the DAOPHOT II[50] PSF-fitting photometry package, with the photometric zero-points taken from the calibration work at STScI.

**Transit search and candidate selection**

As described in the main text, power spectra were calculated for each time series in order to detect the periodic transit signals. To evaluate the limiting S/N at which the number of false positive detections would be expected to be less than one, we carried out a blind experiment using simulated transit signals. About 100 artificial planetary-transits were blindly inserted into a simulated timeseries dataset of 115,000 non-variable stars with identical noise characteristics to the real SWEEPS data. The data were then processed through our software to produce a list of transit candidates and S/N values as described above. All detected transits for which the calculated S/N exceeded 6.5 proved to be genuine events, with the first false positives appearing below this value. Hence a S/N limit of 6.5 was imposed to avoid any statistical false positives in our detected candidates.

**Detection efficiency**

Detection efficiency is defined as the fraction of actual transiting planets in the photometric data that is recovered by our search technique. To estimate this efficiency, ~80,000 transits were introduced blindly into the actual time-series data, including appropriate noise. The simulated transit light curves assumed planet radii of 0.6 to 1.4 $R_J$, a random period distribution over the interval 0.3 to 5.0 days, and an appropriate range of transit durations. Our search technique was then used to produce a list of planetary candidates, including their physical characteristics. The results were compared with the input values to derive the detection efficiencies as a function of planetary radius, stellar magnitude, and orbital period.

As expected, shorter periods are more easily detected than longer ones, and transits of brighter stars are more easily recovered than for fainter stars. At short periods of 0.5 to 1.5 days, the detection efficiency is 90% at V=21 for radii of 1.2 Rj, 50% at V = 22.7, and 10% at V=26. The corresponding stellar mass are 0.95, 0.77, and 0.46 $M_\odot$, respectively. For P = 2.5 to 3.5 days, and R ~ 1.2 $R_J$, the detection efficiency is 90% at V=20, 50% at V = 21.3, and 10% at V=24, corresponding to stellar masses of 1.08, 0.88, and 0.65 $M_\odot$. Even at short periods and a planetary radius of 1.4 $R_J$, the efficiency drops close to zero at V = 26.5, corresponding to a stellar mass of 0.42 $M_\odot$.

There are ~180,000 stars brighter than V=27 that form our targets for transit search. As stated above, the efficiency of planet detection drops to zero at V=26.5, and our faintest candidate host has a brightness of V=26.2.

**Supplementary Information** is linked to the online version of the paper at www.nature.com/nature.

**Acknowledgements** This article is based on observations made with the NASA/ESA Hubble Space Telescope, obtained at the Space Telescope Science Institute, which is operated by the AURA, Inc. under a NASA contract, and with the Very Large Telescope at the ESO, Paranal, Chile. We are grateful to Ron Gilliland for his generous contribution of time and efforts for this project. We thank David Bradstreet, Ian Jordan, Geza Kovacs, Don VandenBerg, and the late Andy Lubenow for their help at various stages of the project. We thank an anonymous referee for very useful suggestions.

**Author Information** Correspondence and requests for materials should be addressed to K.C.S. (e-mail: ksahu@stsci.edu).